\documentclass{aa}
\usepackage[dvips]{graphicx}

\begin{document}

\title{Modeling the interstellar aromatic infrared bands with co-added spectra of PAHs}

\author{Amit Pathak\footnote{Current address: Inorganic and Physical Chemistry Department, I.I.Sc., Bangalore, India.}
\and Shantanu Rastogi}

\offprints{S. Rastogi (shantanu\_r@hotmail.com)}

\institute{Physics Department, D.D.U. Gorakhpur University, Gorakhpur - 273009, INDIA.}

\date{Received / Accepted}

\abstract
{}
{The observed variations in profiles of the interstellar aromatic infrared bands correlate with the object type and are indicative of PAH populations existing in different sources. Spectroscopic studies on PAHs can provide tools for the interpretation of variations accompanying the AIBs. As the observed spectra results from a mix of possible species in the region attempt is made to model this composite spectra by co-adding emissions from PAHs in different size groups.}
{Theoretical IR data of PAHs having 10 to 96 carbon atoms is used to obtain emission spectra. The models are taken in size groups making up of small, medium and large PAHs.}
{The models show good profile match with observations for the 7.7 $\mu m$ complex having sub-features at 7.6 and 7.8 $\mu m$. The 7.6 $\mu m$ sub-feature dominates in the spectra of medium sized PAH cations matching observations from UV rich interstellar environments. The 7.8 $\mu m$ component is more intense in the spectra of large PAH cations (model III) correlating with observations from benign astrophysical regions. A possible interpretation for the observations of $C-H$ out-of-plane bend modes and the weak outliers on the blue side of the intense 11.2 $\mu m$ band is proposed. The models provide pointers to possible PAH populations in different regions.}
{}


\titlerunning{Modeling AIBs with co-added spectra of PAHs}
\authorrunning{A. Pathak \& S. Rastogi}

\maketitle

\section{Introduction}

The mid-infrared emission bands at 3.3, 6.2, 7.7, 8.6, 11.2 and 12.7 $\mu m$ (3030, 1610, 1300, 1160, 890 and 790 $cm^{-1}$) and beyond (Gillett et al. \cite{Gillett73}; Cohen et al. \cite{Cohen89}; Geballe et al. \cite{Geballe89}; ISO results \cite{AA96}; Peeters et al. \cite{Peeters02}; Peeters et al. \cite{Peeters04a}) observed in diverse astrophysical sources are popularly known as aromatic infrared bands (AIBs). These bands have been attributed to emission from isolated gas phase PAH molecules present in the interstellar medium (ISM) (L\'eger \& Puget \cite{Leger84}; Allamandola et al. \cite{ATB85}; Puget \& L\'eger \cite{Puget89}; Allamandola et al. \cite{ATB89}). The ubiquity of AIBs make PAHs an important interstellar family of molecules having significant carbon reserves. The emission process of a PAH molecule involves the absorption of an energetic photon ranging from near UV to near IR (Allamandola et al. \cite{ATB89}; Li \& Draine \cite{Li02}; Smith et al. \cite{Smith04}; Mattioda et al. \cite{Mattioda05a}; Mattioda et al. \cite{Mattioda05b}), which renders it to a vibrationally excited state through internal vibrational redistribution (Allamandola et al. \cite{ATB89}; Puget \& L\'eger \cite{Puget89}). PAH molecules de-excite mainly through IR fluorescence with the emission bands corresponding to fundamental modes of vibrations set up within the molecule. PAHs may also relax via visible phosphorescence and/or fluorescence and are proposed as carriers of the extended red emission (ERE) (Witt et al. \cite{Witt06}) and the blue luminescence (BL) (Vijh et al. \cite{Vijh05}).

Extrapolating the spacing between 6.2 and 7.7 $\mu m$ PAH bands produced from experimental data, Hudgins \& Allamandola (\cite{Hudgins99b}) have concluded that PAHs with 50--80 carbon atoms dominate the mid-IR emission. Studies testing the photo-physical stability of PAHs (Schutte et al. \cite{Schutte93}; Allain et al. \cite{Allain96a}; Allain et al. \cite{Allain96b}; LePage et al. \cite{LePage03}) have reached a similar conclusion. Devoted experimental (Szczepanski \& Vala \cite{Szczepanski93}; Hudgins et al. \cite{Hudgins94}; Hudgins \& Allamandola \cite{Hudgins95}; Hudgins \& Sandford \cite{Hudgins98}; Cook et al. \cite{Cook96}; Kim et al. \cite{Kim01}; Kim \& Saykally \cite{Kim02}) and theoretical research (Langhoff \cite{Langhoff96}; Bauschlicher \& Bakes \cite{Bauschlicher00}; Bauschlicher \cite{Bauschlicher02}; Pathak \cite{Pathak06}; Pathak \& Rastogi \cite{PR05}; Pathak \& Rastogi \cite{PR06}; Pathak \& Rastogi \cite{PR07}) has resulted in considerable IR information of large number of PAHs of varying sizes and ionized states. These results serve as inputs to theoretical emission models (Bakes et al. \cite{Bakes01a}; Bakes et al. \cite{Bakes01b}; Pech et al. \cite{Pech02}; Joblin et al. \cite{Joblin02}; Mulas et al. \cite{Mulas03}; Mulas et al. \cite{Mulas06a}; Mulas et al. \cite{Mulas06b}) for direct comparison with observations. Such modeling studies have emphasized the contribution of PAH cations towards AIBs along with neutral PAHs contributing to a lesser extent. The fractional abundance of cations and neutrals depends on the astrophysical environment where the AIBs are being observed. Current understanding points towards dominance of cations in harsh UV rich conditions of reflection nebulae and HII regions while a mixture of neutrals and cations in relatively cool environments of late type carbon rich stars (Allamandola et al. \cite{AHS99}).

Over the years, developments in observational accuracy using space telescopes have improved the quality of data and a large and diverse sample of astrophysical sources have been examined that exhibit these mid-IR bands (Peeters et al. \cite{Peeters02}; Peeters et al. \cite{Peeters04a}; Van Diedenhoven et al. \cite{vanD04}). Observations from ISO (\cite{AA96}) and SPITZER (\cite{Spitzer}) have provided good quality data that has revealed significant variations and finer profile details of the individual AIBs. Of great interest is the peak position of 6.2 $\mu m$ band, the broad composite 7.7 $\mu m$ band (Peeters et al. \cite{Peeters02}) and variations accompanying the asymmetric 11.2 $\mu m$ band, features at 12.7, 13.3 $\mu m$ and at higher wavelengths (Hudgins \& Allamandola \cite{Hudgins99a}; Hony et al. \cite{Hony01}). Few new features detected around 6.7, 10.1, 15.8, 17.4 and 19.0 $\mu m$ have further added to the AIB numbers (Werner et al. \cite{Werner04}; Peeters et al. \cite{Peeters04b}). These revelations make studies on different PAH groups important to obtain better understanding of AIBs.

In this report we attempt to model the variations associated with AIBs using our existing IR database of PAHs (Pathak \cite{Pathak06}; Pathak \& Rastogi \cite{PR05}; Pathak \& Rastogi \cite{PR06}; Pathak \& Rastogi \cite{PR07}). The IR data consisting of PAHs with 10 to 96 carbon atoms is computed using the GAMESS ab-initio program (Schmidt et al. \cite{Gamess}) under DFT using B3LYP functionals in conjunction with 4-31G basis expansion. To model the observed profile variations in AIBs first the thermal emission model (Schutte et al. \cite{Schutte93}; Pech et al. \cite{Pech02}) is used to obtain the emission spectra of each PAH. Then for PAHs in different size groups the emission spectra is co-added to model AIB spectra. The comparisons with observations give insight into possible PAH size distribution and their processing in different astrophysical regions. Attempt is made to put constraints on the PAH population residing in different interstellar environments.

\section{Emission model}

For a meaningful comparison with observations emission spectra of PAHs is needed (Schutte et al. \cite{Schutte93}; Cook \& Saykally \cite{Cook98}; Pech et al. \cite{Pech02}; Mulas et al. \cite{Mulas03}). L\'eger et al. (\cite{Leger89}) and Schutte et al. (\cite{Schutte93}) have shown that cooling of a PAH via transitions in an emission cascade can be described by the thermal approximation if the average energy of the mode under consideration is small compared to the total energy of the excited PAH. The thermal model is known to break down for emission in low energy modes near the end of the emission cascades (Joblin et al. \cite{Joblin02}; Mulas et al. \cite{Mulas06a}; Mulas et al. \cite{Mulas06b}). In the present work a strong background radiation field, corresponding to blackbody temperature $T = 40,000 K$, is considered. This ensures that the total internal energy of excited PAH is much higher than the energy of any emission mode and thermal approximation holds. A similar approach is taken by Cook \& Saykally (\cite{Cook98}) and Pech et al. (\cite{Pech02}).

We compute the emission spectrum for each individual PAH with calculated absorption spectrum used as input. As in Schutte et al. (\cite{Schutte93}), the PAHs are considered in a exciting radiation field similar to that surrounding $\Theta^1$ Ori C, corresponding to blackbody temperature $T_* = 40,000 K$. The PAH molecules absorb photons of all energies with cut off at 13.6 eV giving rate of photon absorption as:
\begin{equation}
R_{abs} = \int_{0}^{13.6} \frac {B_{\nu}\sigma_{\nu}}{h\nu} d{\nu}
\end{equation}
where $B_{\nu}$ is the Planck function, $\sigma_{\nu}$ is the photo-absorption cross-section and $\nu$ are the absorbed UV radiation frequencies. For each PAH $\sigma_{\nu}$ are taken from \mbox{$http://astrochemistry.ca.astro.it/database/$} (Malloci et al. \cite{Malloci07}). The differences in the shape of the photo-absorption is apparent mostly in the low energy part of the spectrum, which for the UV-rich spectrum in the present emission models is irrelevant. So, for PAHs having no entry in the database, scaled photo-absorption cross-section of PAHs closest in size (number of carbon atoms) is used. The scaling is approximated by a factor $N_{C2}/N_{C1}$ (Mulas et al. \cite{Mulas06b}), with PAH having $N_{C1}$ carbon atoms as proxy for larger molecule with $N_{C2}$ carbon atoms.

The absorption of a high energy photon of frequency $\nu$ excites the PAH molecule to an internal energy equivalent to peak temperature $T_{p}$ that depends on the heat capacity of the PAH. In the harmonic approximation this $T_{p}$ is obtained from:
\begin{equation}
U(T) = \sum_{i=1}^{m} \frac {hc~\omega_{i}}{exp(hc~\omega_{i}/kT) - 1}
\end{equation}
where $i$ corresponds to vibrational modes within individual PAHs having frequency $\omega_{i}$ in $cm^{-1}$ and $m$ is the total number of vibrational modes (3N -- 6; N being the number of atoms).

To ensure the validity of the thermal approximation fall in peak temperature for each PAH is calculated assuming integrated emission of highest frequency modes corresponding to the $C-H$ stretch vibrations. For naphthalene, the smallest PAH in the sample, peak temperature falls by 11.3\%, which is maximum. While for $C_{96}H_{24}$, the largest PAH, fall in peak temperature is less than 0.5\%. This is sufficiently small to justify the use of thermal model in the exciting radiation field considered.

The molecule cools down from the peak temperature $T_p$ through its different vibrational modes with cascade transitions from levels $v \rightarrow v-1$. The emission photon flux $\phi_{i}$ of the $i^{th}$ mode is given as:
\begin{equation}
\phi_{i} = A^{1,0}_{i} \times [\exp(hc~\omega_{i}/kT) -1]^{-1}
\end{equation}
where the Einstein coefficients $A^{1,0}_{i}$ are obtained from the absorption intensities $S_{i}$ (in units of $Km/mol$) (Cook \& Saykally \cite{Cook98}) by the relation:
\begin{equation}
A^{1,0}_{i} = (1.2512 \times 10^{-7}) ~\omega_{i}^{2}~S_{i}
\end{equation}

For a fall in internal energy by $\Delta U$ the fractional energy emitted in the $i^{th}$ mode is then given as:
\begin{equation}
\Delta E_{i}(T) = \frac {\phi_{i} \times \omega_{i}}{\sum_{i=1}^{m} \phi_{i} \times \omega_{i}} \times \Delta U(T)
\end{equation}
The fractional energy $E_{i}$ is integrated over the cooling range from $T_p$ to a temperature of 50 K below which the energy emitted is negligible. $\Delta U$ corresponding to fall in temperature by 1 K is taken at each $T$. This emitted energy is weighted by the rate of photon absorption (Eq. 1) and integrated over the whole distribution of absorbed photons.

The emission spectra thus obtained for each PAH is then used to model composite spectrum from groups of PAHs. Spectra of PAHs in each group are co-added assuming equal number of each specie in the ISM.

\section{Spectral models and their astrophysical significance}

\begin{figure}[!h]
\resizebox{\hsize}{!}{\includegraphics{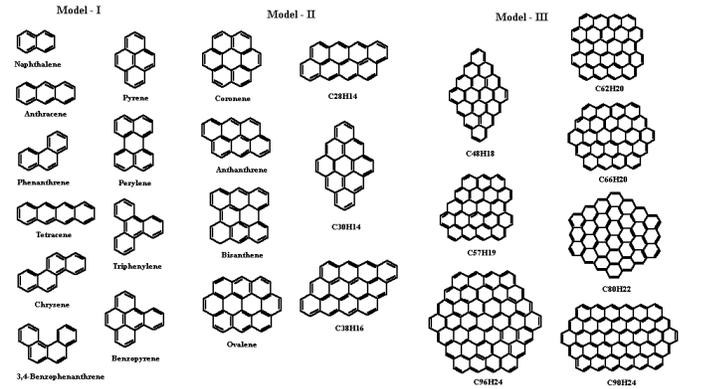}}
\caption{PAHs constituting the three models.}
\label{Fig1}
\end{figure}

AIBs result from composite emission of a number of PAHs consequently profile variations associated with AIBs in different sources reflect the presence of different PAH groups therein. Factors like anharmonicity and rotational broadening also affect line widths and shape (Pech et al. \cite{Pech02}). Anharmonicity induces small asymmetry in the profile and rotational widths too are small and get smaller for larger PAHs. Contribution of these effects is $< 5$ $cm^{-1}$ (Mulas et al. \cite{Mulas06a}; Pech et al. \cite{Pech02}), while typical peak width under interstellar conditions is $\sim$ 30 $cm^{-1}$ (Allamandola et al. \cite{ATB89}). Experimental measurements on a few individual PAHs show a near linear dependence on band width and temperature with intrinsic widths between 10 - 20 $cm^{-1}$ around 1000 K (Joblin et al. \cite{Joblin95}; Pech et al. \cite{Pech02}). In obtaining the composite emission spectra we neglect anharmonicity and rotational broadening and assume a Lorentzian profile with FWHM of 20 $cm^{-1}$. Several modes closer than this may collocate together and result in a single broad peak. Three distinct models are presented to understand the gross changes (intensity as well as profile variations) in the PAH spectra with different PAH size groups. Individual PAHs used in the models are displayed in Fig. \ref{Fig1}.

\begin{figure}[!h]
\resizebox{\hsize}{!}{\includegraphics{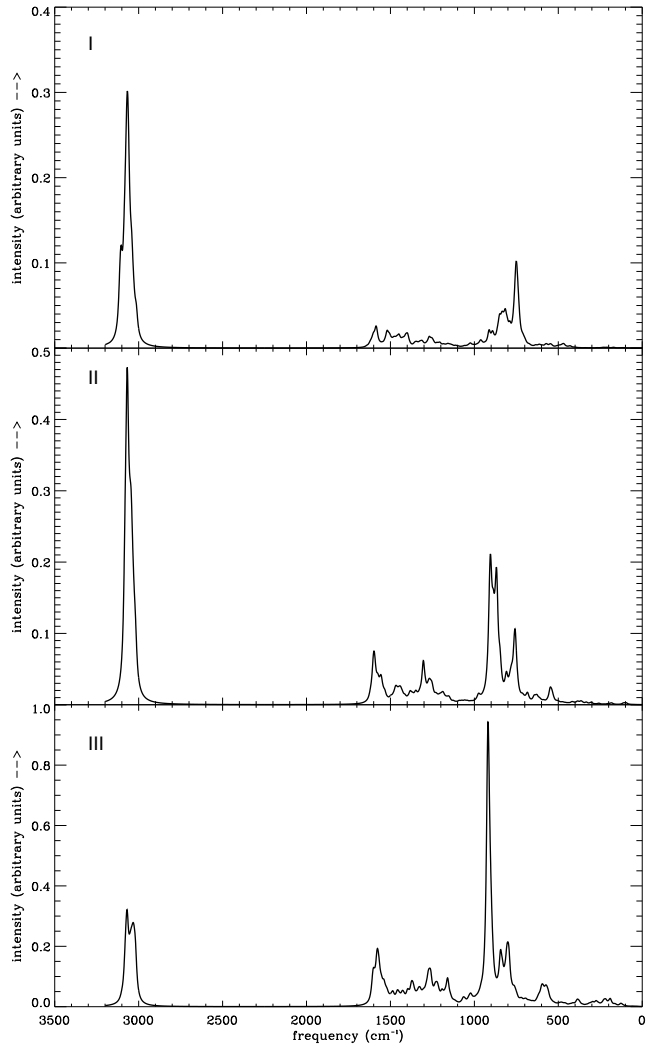}}
\caption{Model spectra of co-added emissions from neutral PAHs with (I) less than 20 C atoms, (II) 20 to 40 C atoms, (III) more than 40 C atoms.}
\label{Fig2}
\end{figure}

Model I comprises of co-added emission spectra of small PAHs ($\leq 20$ carbon atoms) consisting of naphthalene, anthracene, phenanthrene, pyrene, tetracene, chrysene, 3,4-benzophenanthrene, triphenylene, perylene and benzopyrene. Model II (PAHs with 20 to 40 carbon atoms) is derived from a mixture of anthanthrene, coronene, bisanthene, ovalene, $C_{28}H_{14}$, $C_{30}H_{14}$ and $C_{38}H_{16}$. Model III uses the co-added spectra of large PAHs ($>$ 40 carbon atoms) which includes $C_{48}H_{18}$, $C_{57}H_{19}$, $C_{62}H_{20}$, $C_{66}H_{20}$, $C_{80}H_{22}$, $C_{90}H_{24}$ and $C_{96}H_{24}$. All the three models consider equal number of constituent PAHs. Emission spectra of the three models as neutral and cation are compared in Fig. \ref{Fig2} and Fig. \ref{Fig3}. In the composite spectra each peak may result from the overlap of different modes in different PAHs. Therefore only a general assignment of intensity peaks with vibrational modes is possible.

The spectra of neutrals (Fig. \ref{Fig2}) is dominated by the $C-H$ stretch and $C-H$ out-of-plane bend vibrations. Comparison of the absorption calculations with experimental data wherever available shows that the $C-H$ stretch intensity is overestimated by about 1.4 to 2.0 times (Pathak \& Rastogi \cite{PR07}). Use of larger basis sets, i.e., 6-31G or 6-31G**, in place of 4-31G, provide better representation of the orbitals and consequently a good $C-H$ stretch intensity match with experiments. Use of larger basis for some of the PAHs incorporated in model I shows that computational effort increases greatly and the scaling procedure also gets complicated but there is negligible change in intensity or position of modes other than $C-H$ stretch (Pathak \& Rastogi \cite{PR07}). Therefore, for the study of profiles and relative intensity of all other modes the 4-31G basis is suitable while giving qualitative information of the $C-H$ stretch.

In the cascade emission model low temperatures contribute little towards the intensity of modes at higher frequencies. The contribution is diminished further in case of large PAHs that are excited to lower peak temperatures. In large PAH model III the intensity of $C-H$ stretch mode is small compared to the models with smaller PAHs. The observed $C-H$ stretch (3.3 $\mu m$) and $C-H$ out-of-plane bend (11.2 $\mu m$) intensity ratio is a suitable parameter constraining the size distribution of PAHs (Schutte et al. \cite{Schutte93}; Pech et al. \cite{Pech02}).

The absorption intensity of $C-H$ stretch vibrations reduce drastically upon ionization while small intensity variations are observed for the $C-H$ wag modes. Intensity of $C-H$ stretch mode depends on the charge density near the hydrogen atoms (Hudgins et al. \cite{Hudgins01}; Pathak \& Rastogi \cite{PR05}; Pathak \& Rastogi \cite{PR06}). In cations this (positive) charge density increases consequently the $C-H$ stretch intensity gets drastically reduced. In larger PAH cations due to the distribution of the acquired positive charge over more number of atoms the absorption in $C-H$ stretch mode is significant (Pathak \& Rastogi \cite{PR06}; Pathak \& Rastogi \cite{PR07}). This and the temperature effect in cascade emission model is reflected in the composite emission spectra (Fig. \ref{Fig3}) showing significant $C-H$ intensity in model II and reduced intensity in model I and model III.

\subsection{$C-C$ stretch modes}

Among the AIBs the 6.2 and 7.7 $\mu m$ emission features are most intense. The $C-C$ stretch vibrations set up in ionized PAH molecules give rise to these bands. The 6.2 $\mu m$ band has some contributions from $C-H$ in plane bend modes as well (Pathak \& Rastogi \cite{PR05}). Profile variations observed in these bands are a direct measure of the background environments that excite the PAHs.

Peeters et al. (\cite{Peeters02}) report detailed ISO observations of 57 diverse sources in the 6 to 9 $\mu m$ range and based on intricate position and profile variations classify these bands into three classes $A$, $B$ and $C$. Class $A$ has the 6.2 $\mu m$ band maximum in between 6.20 and 6.23 $\mu m$ and the 7.7 $\mu m$ complex peak at 7.6 $\mu m$. Bands between 6.23 and 6.29 $\mu m$ and a peak at 7.8 $\mu m$ are classified as class $B$. Bands centered at 6.29 and redder are classified as class $C$. Out of the 57 sources reported, 42 belong to class $A$, 12 to class $B$ and 2 sources are classified as $C$. Similar profiles have also been reported by other observations for reflection nebulae and Herbig Ae/Be stars (Bregman \& Temi \cite{Bregman05}; Sloan et al. \cite{Sloan05}).

\begin{figure}[!h]
\resizebox{\hsize}{!}{\includegraphics{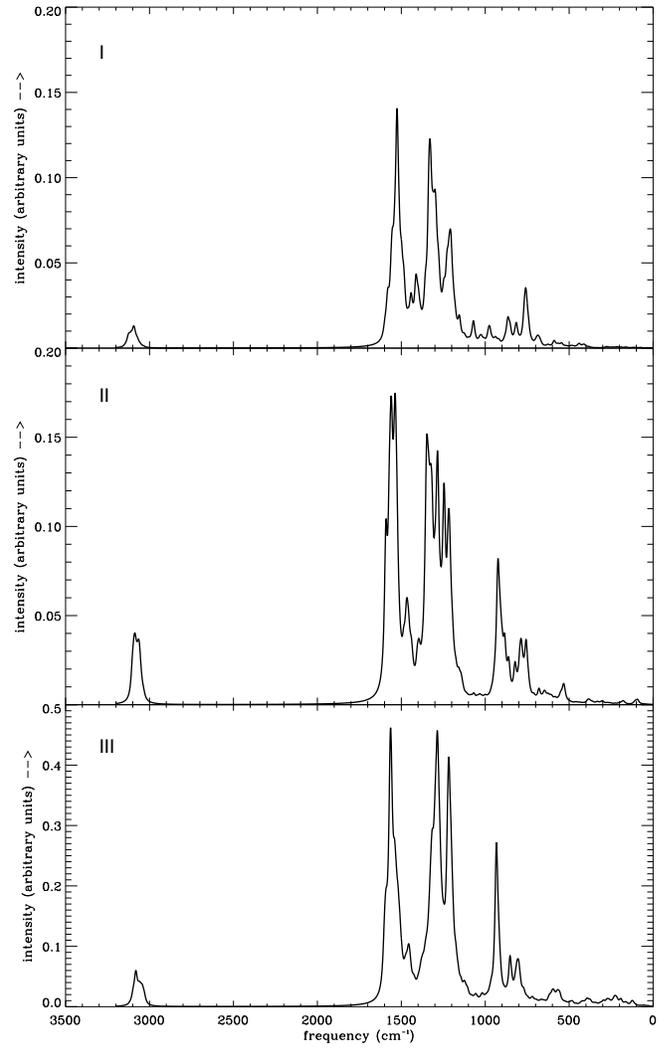}}
\caption{Model spectra of co-added emissions from PAH cations with (I) less than 20 C atoms, (II) 20 to 40 C atoms, (III) more than 40 C atoms.}
\label{Fig3}
\end{figure}

All three PAH cation models (Fig. \ref{Fig3}) show intense broad peaks corresponding to 7.7 and 6.2 $\mu m$ modes. The peaks are at 1330 (7.52 $\mu m$) and 1526 $cm^{-1}$ (6.55 $\mu m$), 1347 (7.42 $\mu m$) and 1537 $cm^{-1}$ (6.50 $\mu m$) and at 1285 (7.78 $\mu m$) and 1564 $cm^{-1}$ (6.39 $\mu m$) respectively for models I, II and III. The $C-H$ in-plane vibrations have peaks around 1215 $cm^{-1}$ (8.23 $\mu m$) in all the models with little variation in the peak position.

Observations show that the 7.7 $\mu m$ AIB incorporates two sub-components, at 7.6 $\mu m$ (1315 $cm^{-1}$) and at 7.8 $\mu m$ (1282 $cm^{-1}$) (Cohen et al. \cite{Cohen89}; Bregman et al. \cite{Bregman89}; Peeters et al. \cite{Peeters02}). The 7.6 $\mu m$ feature tends to dominate the emission spectra of regions involved in processing of PAHs and other molecules i.e., star forming regions, HII regions and reflection nebulae, etc. The 7.8 $\mu m$ component dominates in environments where the PAHs are relatively fresh and unprocessed for example, planetary nebulae and objects evolving from Asymptotic Giant Branch (AGB) phase of stellar evolution. Similar sub-components for the 7.7 $\mu m$ band are obtained in the present models. This provides useful insight regarding the PAH size and charge state in different interstellar environments.

\begin{figure}[!h]
\includegraphics[height=8in]{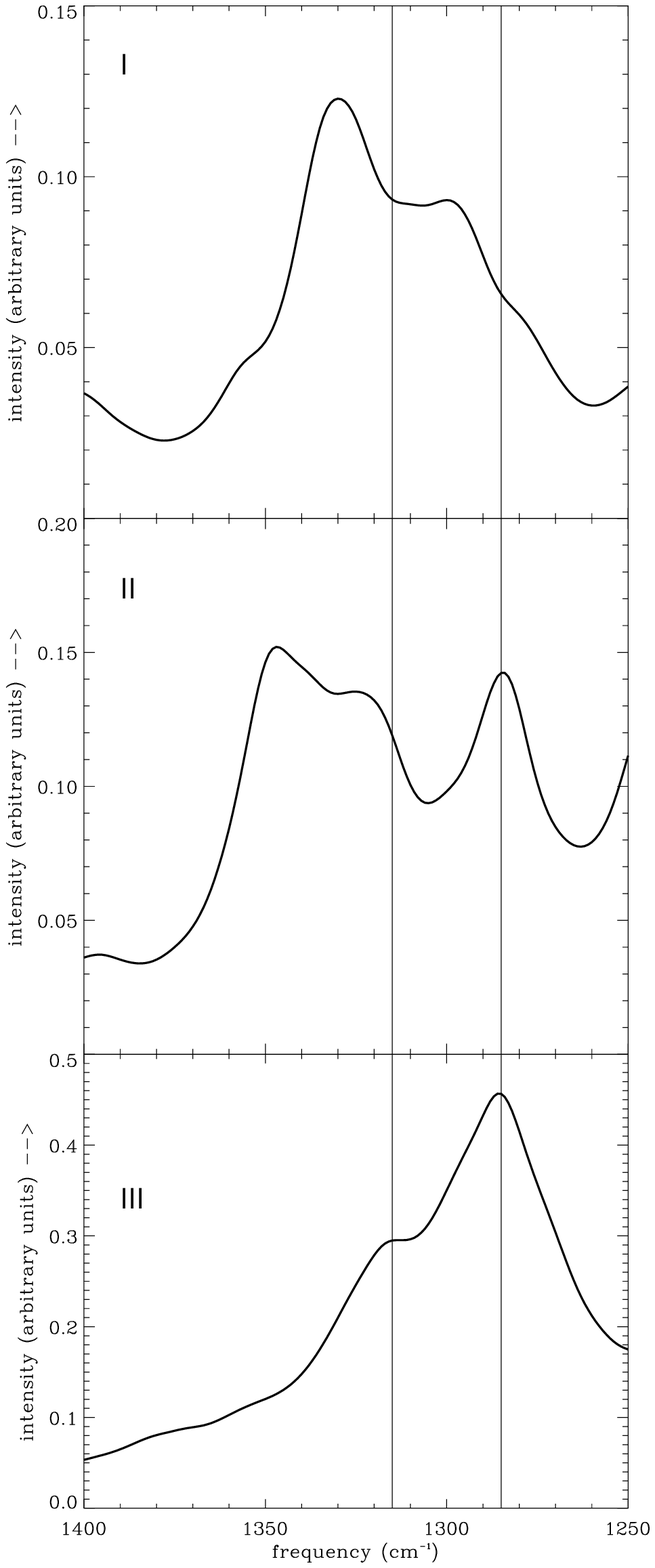}
\caption{Emission spectra of three cation models in the 1250 to 1400 $cm^{-1}$ region. The observed 7.7 $\mu m$ AIB lies within the two parallel lines.}
\label{Fig4}
\end{figure}

Fig. \ref{Fig4} shows the expanded spectra, in the 1250 -- 1400 $cm^{-1}$ region, for the cation models of Fig. \ref{Fig3}. The two parallel lines at 1285 and 1315 $cm^{-1}$ enclose the observed 7.7 $\mu m$ complex. Two sub-features which constitute the 7.7 $\mu m$ band having a separation of 30 -- 40 $cm^{-1}$ are evident in the three models. Model I has a dominant feature at 1330 $cm^{-1}$ (7.52 $\mu m$) corresponding to the observed 7.6 $\mu m$ component along with a shoulder at 1300 $cm^{-1}$ (7.7 $\mu m$). In model II the lower wavelength component around 1347 $cm^{-1}$ (7.42 $\mu m$), with shoulder at 1325 $cm^{-1}$ (7.55 $\mu m$), dominates while the higher wavelength sub-feature at 1285 $cm^{-1}$ (7.78 $\mu m$) is less intense. Profiles similar to model I and II, with stronger lower wavelength component, have been observed in UV-rich environments of HII regions and reflection nebulae and have been classified as $A^\prime$ profiles (Fig. 13 in Peeters et al. \cite{Peeters02}). Medium sized compact PAHs with 40 -- 50 carbon atoms, as present in model II, are quite stable and may survive in strong UV environments providing good representation of PAHs populating such interstellar sources. Observations also indicate that sources dominated by the 7.6 $\mu m$ sub-feature always exhibit a distinct 7.8 $\mu m$ sub-feature (Peeters et al. \cite{Peeters02}). Model II fulfills this observational norm well.

In model III that comprises large PAHs, the 7.7 $\mu m$ band is dominated by the higher wavelength sub-component at 1285 $cm^{-1}$ (7.78 $\mu m$) with an apparent shoulder around 1315 $cm^{-1}$ (7.60 $\mu m$). Unlike model II it is not distinctly resolved but appears as a shoulder to the main peak. Observations show that sources with dominant 7.8 $\mu m$ component do not always have a very clear and distinctive 7.6 $\mu m$ component (Peeters et al. \cite{Peeters02}; Bregman \& Temi \cite{Bregman05}). The profile of model III conforms with these observations of relatively benign astrophysical regions classified as $B^\prime$ profiles (Peeters et al. \cite{Peeters02}).

For direct comparison and matching with observations, intensity ratios of the two components of the 7.7 $\mu m$ complex  and their wavenumber separation in the three models is presented in Table \ref{table1}. Similar ratios from observations by Peeters et al. (\cite{Peeters02}) have also been provided. A good match in terms of intensity ratio is present with different observed sources. The profile observed in model I is similar to the classification $A^{\prime}$. Model II shows a separation of almost 62 $cm^{-1}$ between the two components of the composite 7.7 $\mu m$ band with an intensity ratio of 1.06. The separation is a bit more than that is observed for $A^{\prime}$ profiles. The intensity ratio is also less than what is measured for NGC 2023 but is closer to that in IRAS 23133. Similar match has been shown to exist with the co-added spectra of medium sized pericondensed PAH cations (Pathak \& Rastogi \cite{PR06}). Model III shows a clear agreement with the observed $B^\prime$ profiles. The separation between the two components of the 7.7 $\mu m$ band as well as the intensity ratio match well especially for NGC 7027.

\begin{figure}[!h]
\includegraphics[height=8in]{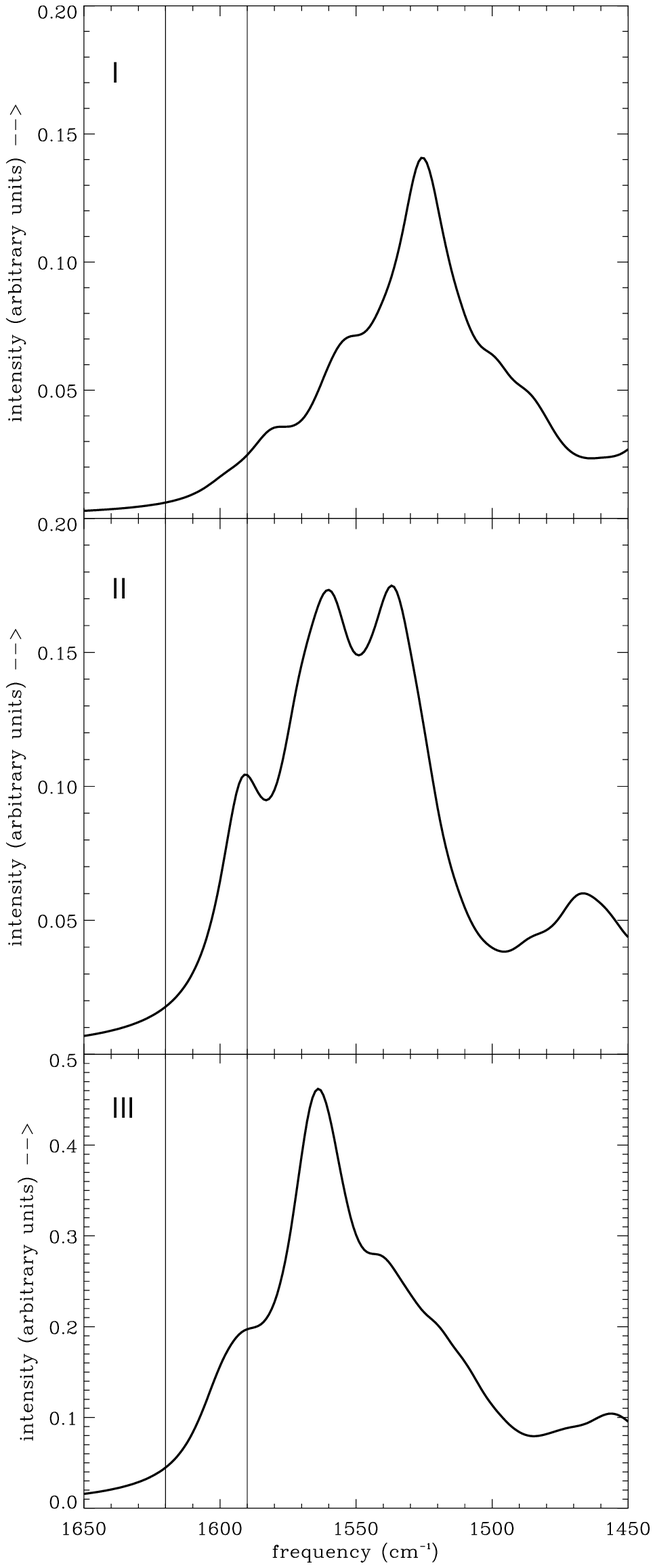}
\caption{Emission spectra of three cation models in the 1450 to 1650 $cm^{-1}$ region. The observed 6.2 $\mu m$ AIB lies within the two parallel lines.}
\label{Fig5}
\end{figure}

The region of 6.2 $\mu m$ feature, i.e. 1450 to 1650 $cm^{-1}$ region of Fig. \ref{Fig3}, is expanded in Fig. \ref{Fig5}. There is no position match with the 6.2 $\mu m$ (1610 $cm^{-1}$) AIB in any of the three models. The co-added spectra of small PAHs (model I) has this band centered at 1526 $cm^{-1}$ (6.55 $\mu m$) which is far off from the observed 6.2 $\mu m$ band. As the PAH size increases a blue shift, similar to available experimental data (Hudgins \& Allamandola \cite{Hudgins99b}), is observed. The co-added spectra of medium sized PAHs (model II) has this feature at 1537 $cm^{-1}$ (6.51 $\mu m$) and for large PAHs (model III) only a small additional shift to 1564 $cm^{-1}$ (6.39 $\mu m$) is present. Few PAH cations eg., $C_{28}H_{14}$, $C_{38}H_{16}$ and $C_{90}H_{24}$ (Pathak \& Rastogi \cite{PR06}; Pathak \& Rastogi \cite{PR07}) independently show intense feature around 6.29 $\mu m$ (1590 $cm^{-1}$) corresponding to the observed class $C$ band (Peeters et al., \cite{Peeters02}). These PAHs contribute to the small shoulder observed beyond 1590 $cm^{-1}$ (6.29 $\mu m$).

Failure to explain the position of the 6.2 $\mu m$ AIB with co-added model spectra of pure PAHs points towards other possibilities. Recent theoretical calculations on nitrogen substituted large PAH cations (PANHs) proposes to explain the variations in the peak position of the 6.2 $\mu m$ interstellar emission feature (Hudgins \& Allamandola \cite{Hudgins03}; Hudgins et al. \cite{Hudgins05}). Incorporating nitrogen inside the ring, i.e., replacing one or more carbon atoms results in a shift of the band from 6.4 $\mu m$ to 6.2 $\mu m$. Similar blue shift is also reported for Iron -- PAH coordination complex (Simon \& Joblin \cite{Simon07}). Hydrogenated, de-hydrogenated and multiply charged PAHs, PAH clusters, anions and substituted PAHs (Bakes et al. \cite{Bakes01a}, Halasinski et al. \cite{Halasinski05}, Malloci et al. \cite{Malloci05}, Malloci et al. \cite{Malloci07}, Vuong \& Foing \cite{Vuong00}) also need careful attention and they cannot be completely neglected. Either these or a combination of these with pure PAH ions may be responsible for the 6.2 $\mu m$ emission feature.

\subsection{$C-H$ out-of-plane bend modes}

The 11.2 $\mu m$ (893 $cm^{-1}$) AIB represents solo out-of-plane wag modes in PAHs. The weak 12.7 $\mu m$ (787 $cm^{-1}$) band and features at longer wavelengths represent trio and quartet hydrogen wag modes. ISO observations by Hony et al. (\cite{Hony01}) conclude that the spectra of planetary nebulae and evolved carbon-rich stars, where PAHs are supposed to be synthesized and are still unprocessed, have a strong 11.2 $\mu m$ feature, weak 12.7 $\mu m$ band and features at longer wavelengths. The situation is almost reversed in case of UV-rich environments where the 12.7 $\mu m$ band is as intense as the 11.2 $\mu m$ band.

\begin{figure}[!h]
\resizebox{\hsize}{!}{\includegraphics{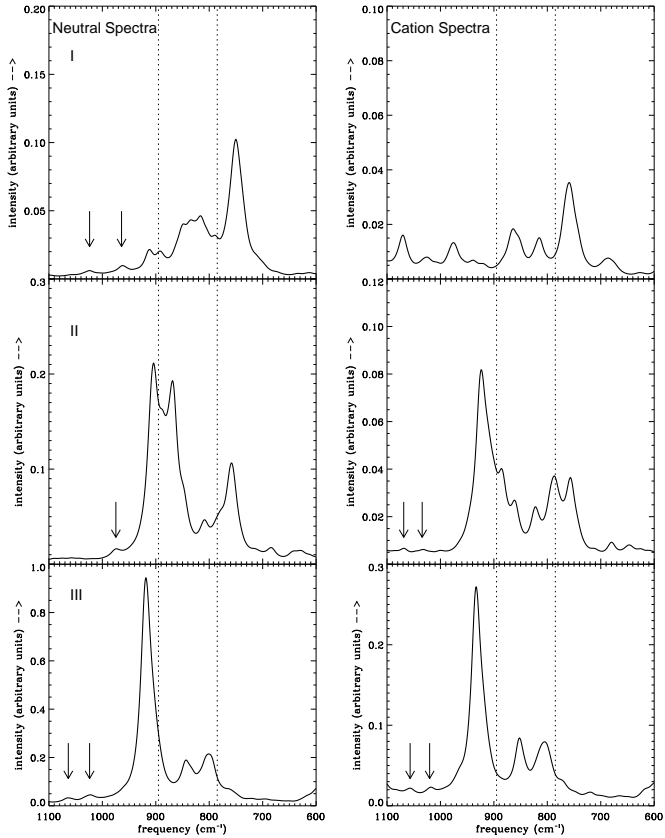}}
\caption{Emission spectra of three models in the 600 to 1100 $cm^{-1}$ region. Arrows point to blue outliers of 11.2 $\mu m$ band. Dotted vertical lines represent central peak of 11.2 and 12.7 $\mu m$ AIB.}
\label{Fig6}
\end{figure}

The spectra of small (model I) and medium sized (model II) neutral PAHs (Fig. \ref{Fig6}) have strong quartet hydrogen wag modes peaking around 750 $cm^{-1}$. Mixing of duo and trio hydrogen wags result in a broad peak around 815 $cm^{-1}$ in small PAHs. For PAHs comprising 20 to 40 carbon atoms (model II) the intense solo wag mode near 900 $cm^{-1}$ (11.11 $\mu m$) shows best position and intensity agreement with the observed 11.2 $\mu m$ AIB. The spectrum of model III has this feature blue-shifted to around 915 $cm^{-1}$ (10.93 $\mu m$). Model III also displays discrete less intense features due to duo and trio hydrogen out-of-plane bend vibrations at 840 and 800 $cm^{-1}$ respectively.

Upon ionization the $C-H$ out-of-plane bend modes in all the PAHs show significant shift in band positions but intensity variations are small (Fig. \ref{Fig6}). Model I has intense peak at 760 $cm^{-1}$ (13.16 $\mu m$) and less intense feature at 864 $cm^{-1}$ (11.57 $\mu m$). These features in model I correlate with observations of weak bands present on the long wavelength side of the 11.2 $\mu m$ band (Witteborn et al. \cite{Witteborn89}). The co-added emission spectra of medium sized (model II) PAH cations has an intense feature at 920 $cm^{-1}$ and significant  peaks at 755 and 785 $cm^{-1}$. Large PAH cations (model III) have prominent solo hydrogen ($C-H$) out-of-plane bend modes producing an intense peak at 933 $cm^{-1}$ and weak features between 800 and 850 $cm^{-1}$. The shifts in band positions of hydrogen out-of-plane modes with ionization and increasing PAH size are apparent in the three models.

Sloan et al. (\cite{Sloan99}) observed complex less intense structures on the short wavelength side of the principal 11.2 $\mu m$ AIB. They referred them as ``blue outliers''. These blue outliers were more noticeable near the central exciting object. The 11.2 $\mu m$ feature adopts a smoother profile in observations away from the central source with weak or no blue outliers. These observations correlate with our models. Weak features on the blue side of the 11.2 $\mu m$ have been pointed by arrows in Fig. \ref{Fig6}. Existence of stable compact medium sized PAHs or cations in the vicinity of the central source seem to be a proposition. This also correlates with the observation that UV rich environments where 6.2 and 7.7 $\mu m$ show $A$ type profile also have $A$ type 11.2 $\mu m$ (Van Diedenhoven et al. \cite{vanD04}).

\subsection{Intensity Ratios}

The intensity of the AIBs may not be directly correlated with spectral models based on theoretical or experimental measurements but the ratios of different features prove to be highly useful. Systematic variations associated with intensity ratios for different modes may be used to constrain PAH size, structure and ionization state in the ISM. In neutral PAHs, where the spectra is dominated by $C-H$ stretch and wag modes with weak $C-C$ stretch mode intensities, the 6.2 and 7.7 $\mu m$ to 11.2 $\mu m$ band ratio is small but for cations tremendous increase in the value is present, which is an indicator of the ionization state of PAHs. The emission model intensity ratios of few features are presented in Table \ref{table2}. Since the models fail to reproduce the 6.2 $\mu m$ feature the presented ratios are not directly relevant for observations but point towards a trend.

For cations, the value of 6.2 to 7.7 $\mu m$ band ratio is smallest for large PAHs. The ratio of 11.2 to 12.7 $\mu m$ PAH bands for neutral PAHs sheds light on the shape (peripheral structure) of PAHs (Hudgins \& Allamandola \cite{Hudgins99a}; Hony et al. \cite{Hony01}). As the PAH size increases, the ratio appears to increase steadily. Since, small PAHs inevitably have more corners, leading to more duo and trio hydrogens, the intensity of their wag is strong. Compact and stable large PAHs considered in the models have straight edges resulting in increased number of solo hydrogens and the spectra has an intense 11.2 $\mu m$ band.

\subsection{PAH formation and processing}

The correlation of observations with model spectra, from collective emission of PAHs, presented in this report provides leads to understanding the formation and processing of PAHs in the ISM. The correlation of 7.7 $\mu m$ $B^\prime$ profile with model III suggests formation of large PAHs (having around 100 carbon atoms) in outflows of late-type carbon rich stars. Since, ionization energy for large PAHs is less, they are more likely to be cations. The spectra of these large PAH cations dominate benign regions of the ISM. Observations of the 6.3 $\mu m$ (1585 $cm^{-1}$) AIB ($C$ class; Peeters et al., \cite{Peeters02}) in these regions suggests that the PAHs initially formed are pure and unsubstituted. The processing of these large systems starts as soon as they are formed. Ruled by the ISM conditions equilibrium is established between formation and destruction. A likely scenario is that in harsh ISM conditions breaking and reduction of large structures along with multiple ionization of medium sized PAHs dominates. Observation of $A^{\prime}$ profiles in UV-rich environments of ISM suggests the abundance of compact medium sized PAH cations. This is also confirmed by the observation of a strong 12.7 $\mu m$ band in UV-rich environments. Formation of multiply charged PAHs, substituted PAHs (PANHs), PAH complexes and clusters is suggested by the observation of 6.2 $\mu m$ $A$ class profile in harsh strong UV flux environments (Peeters et al. \cite{Peeters02}).

Taking cue from these models more specific modeling for particular observations can be made by varying the concentrations of the individual PAHs. These can provide better understanding of the chemistry of the region. A combination of PAHs from model I and II with different concentrations is taken to compare with observed 7.7 $\mu m$ feature of reflection nebula NGC 2023. The most suitable combination obtained has Perylene(14\%), Triphenylene(11\%), Benzo-pyrene(9\%), Ovalene(25\%), Bisanthene(9\%), $C_{30}H_{14}$(17\%), $C_{38}H_{16}$(15\%) and is shown in Fig. \ref{Fig7}(a). The lower wavelength component dominates the 7.7 $\mu m$ band while the higher wavelength sub-feature is slightly less intense. These have been classified as $A^\prime$ profiles (Peeters et al. \cite{Peeters02}) observed in UV-rich environments of HII regions, reflection nebulae and other star forming regions. Starburst galaxies, where there is copious star formation activity, also have similar profile (e.g. average spectra of 11 Starbursts Fig.1 Lutz et al. \cite{Lutz98}). The 7.7 $\mu m$ band in galaxies is useful in the study of AGN--Starburst connection (Schweitzer et al. \cite{Schweitzer06}; Tran et al. \cite{Tran01}; Lutz et al. \cite{Lutz98}).

Fig. \ref{Fig7}(b) compares the 7.7 $\mu m$ feature of model III with post-AGB star HD 44179 that is classified as $B^\prime$ profile (Peeters et al. \cite{Peeters02}) of relatively benign astrophysical regions where PAHs are supposed to be synthesized. A slight shift in the band position and an additional feature near 1215 $cm^{-1}$ (8.2 $\mu m$) mark the difference in agreement with observation.

It is interesting to compare the models with observation over the whole range from 5 -- 15 $\mu m$, as shown in Fig. \ref{Fig8}. Though there is gross similarity, in both cases it is seen that apart from the 7.7 $\mu m$ band other prominent features are either shifted or have intensity mismatch. The 6.2 $\mu m$ band lies at a higher wavelength while the 11.2 $\mu m$ band is blue shifted in the model spectra. Considering PAH population in ISM to be even partially similar to the present models, observations should show the strong features of the models e.g. around 6.5 and 10.8 $\mu m$.

\section{Conclusions}

AIBs are ubiquitously observed in diverse astrophysical sources with their origin commonly ascribed to interstellar PAHs. Recent space based observations have enabled high quality infrared data which has introduced variations in profiles of the observed mid-IR spectra. These variations correlate with the object type and are indicative of PAH populations existing in different sources. To gain insight into the profile variations of PAH bands and for their better interpretation three distinct models using PAHs of different size groups are presented. Correlation of these models with observations throws light on the formation and processing of PAHs in different astrophysical environments.

\begin{figure}[!h]
\resizebox{\hsize}{!}{\includegraphics{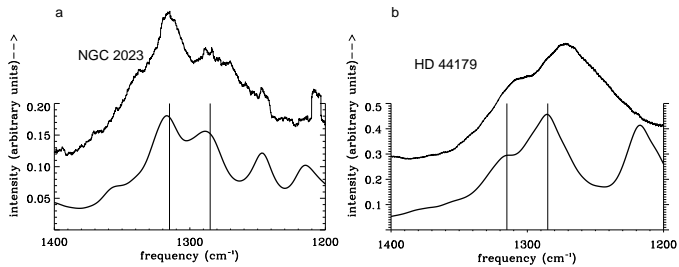}}
\caption{PAH cations emission model in the 1200 to 1400 $cm^{-1}$ region compared with observed spectra (a) PAHs with less than 40 C atoms in varying proportions (see text); (b) more than 40 C atoms (model III). The 7.7 $\mu m$ band lies within the two vertical lines.}
\label{Fig7}
\end{figure}

\begin{figure}[!t]
\resizebox{\hsize}{!}{\includegraphics{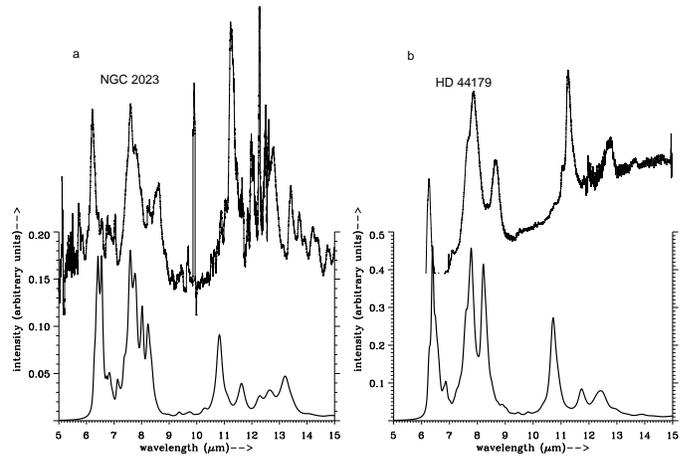}}
\caption{Same as Fig. 7 in the range 5 -- 15 $\mu m$. (a) PAHs with less than 40 C atoms in varying proportions (see text); (b) more than 40 C atoms (model III).}
\label{Fig8}
\end{figure}

The models show good profile match with observations for the 7.7 $\mu m$ band. This feature is not a single peak but a combination of two sub-components at 7.6 and 7.8 $\mu m$ that are also distinct in the presented models. The 7.6 $\mu m$ sub-feature is found to dominate in the spectra of small to medium sized PAH cations matching with the observations of UV rich interstellar environments (Fig. \ref{Fig7}(a)). The 7.8 $\mu m$ component is more intense in the spectra of large PAH cations correlating with the observations of benign astrophysical regions (Fig. \ref{Fig7}(b)). This clearly indicates the formation of large PAHs in the outflows of carbon rich stars that transform to medium sized ones upon processing.

The models also correlate well with the observations of the $C-H$ out-of-plane bend modes and the weak outliers on the blue side of the intense 11.2 $\mu m$ band. Variation in the observed intensity of 12.7 and 11.2 $\mu m$ bands points towards large PAHs in planetary nebula and medium and small PAHs in UV rich environments of HII regions and reflection nebula.

Varying concentrations of the individual PAHs and considering the strength of exciting radiation may allow modeling of specific astrophysical objects. But the feature corresponding to 6.2 $\mu m$ AIB is not correlated with the present set of PAHs. Also strong features of the models, around 6.5, 8.2 and 10.8 $\mu m$, should show up in observations if the PAH population in ISM is similar to those considered here. The models presented do provide pointers to possible PAH populations in different regions, yet for a complete and realistic model same group of PAHs should provide match for all the AIBs.

Failure to explain the complete profile of AIBs may be due to the incompleteness of the sample. A bigger sample of PAHs that includes substituted PAHs, PAH coordination complexes, hydrogenated and de-hydrogenated PAHs and different charge and ionization states of all these may be required. Since laboratory data for most such systems may be difficult to obtain quantum chemical approach is most suitable, which needs to be refined by using larger basis sets. A better modeling approach should incorporate anharmonicity and hot band shifts. Some work has been done in this direction (Joblin et al. \cite{Joblin95}; Pech et al. \cite{Pech02}; Mulas et al. \cite{Mulas06a}; Mulas et al. \cite{Mulas06b}; Can\'e et al. \cite{Cane07}) but exact identification of individual PAHs based on these is still difficult.

\begin{acknowledgements}
The authors are thankful to Prof. Mulas, the referee, for insightful suggestions and improvement in the paper content. The use of High Performance Computing and library facilities at Inter University Center for Astronomy and Astrophysics, Pune is acknowledged.
\end{acknowledgements}

\newpage

\begin{table}
\caption{Band position and strength ratios of the two components of 7.7 $\mu m$ composite.}
\label{table1}
%\centering
\begin{tabular}{l c c c c c l}
\hline\hline
Model     & 7.6  & 7.8  & Diff ($cm^{-1}$)$^1$ & Diff$_{obs}^2$ ($cm^{-1}$) & $I_{7.6}$/$I_{7.8}$$^1$ & Obs I$_{7.6}$/I$_{7.8}$$^2$ \\
\hline
Model I   & 1330 & 1300 & 30 &          & 1.32 & \\
Model II  & 1347 & 1285 & 62 & $\sim$30 & 1.06 & 1.56 (NGC 2023)\\
          &      &      &    &          &      & 1.35 (Orion peak 2) \\
          &      &      &    &          &      & 1.20 (IRAS 23133) \\
Model III & 1315 & 1285 & 30 & $\sim$45 & 0.65 & 0.64 (NGC 7027) \\
          &      &      &    &          &      & 0.52 (IRAS 17047) \\
          &      &      &    &          &      & 0.42 (HD 44179) \\
\hline
\end{tabular}
$^1${\footnotesize Difference of the band position and intensity ratio of the two components of the 7.7 $\mu m$ composite as calculated from the theoretical spectral models.};\\
$^2${\footnotesize The observed results are from Peeters et al. \cite{Peeters02}, Table 2.}
\end{table}

\begin{table}
\caption{Intensity ratios$^{a}$ for the main PAH features.}
\label{table2}
\begin{tabular}{l c c c c}
\hline\hline
Model     & 6.2/7.7   & 6.2/11.2  & 7.7/11.2  & 11.2/12.7 \\
          & (PAH$^+$) & (PAH$^+$) & (PAH$^+$) & (Neutral PAH) \\
\hline
Model I   & 1.15      & 7.65      & 6.66      & 0.44 \\
& & & &\\
Model II  & 1.22      & 2.12      & 1.74      & 2.00 \\
& & & &\\
Model III & 1.01      & 1.70      & 1.68      & 4.39 \\

\hline
\end{tabular}

$a$~--{\footnotesize Ratios are for emission models as in Fig. \ref{Fig2} and Fig. \ref{Fig3}}\\
\end{table}

\newpage

\end{document}